\documentclass[twocolumn,showpacs,preprintnumbers,amsmath,amssymb]{revtex4}
\usepackage{graphicx}
\usepackage{dcolumn}
\usepackage{bm}

\begin{document}
\title{Quantum analysis of shot noise suppression in a series of tunnel 
barriers}
\author {P.~Marconcini}
\author{M.~Macucci} 
\altaffiliation[Also at ]{CNR-IEIIT (Pisa), Via Caruso 16, I-56122 Pisa, Italy}
\author{G.~Iannaccone}
\altaffiliation[Also at ]{CNR-IEIIT (Pisa), Via Caruso 16, I-56122 Pisa, Italy}
\author{B.~Pellegrini}
\altaffiliation[Also at ]{CNR-IEIIT (Pisa), Via Caruso 16, I-56122 Pisa, Italy}
\affiliation{Dipartimento di Ingegneria dell'Informazione,
Universit\`a di Pisa\\
Via Caruso 16, I-56122 Pisa, Italy}
\date{\today}
\begin{abstract}
We report the results of an analysis, based on a straightforward 
quantum-mechanical
model, of shot noise suppression in a structure containing cascaded tunneling
barriers. Our results exhibit a behavior that is in sharp contrast with
existing semiclassical models for this particular type of structure, which
predict a limit of 1/3 for the Fano factor
as the number of barriers is increased. The origin of this discrepancy is
investigated and attributed to the presence of localization on the length
scale of the mean free path, as a consequence of the strictly 1-dimensional 
nature of disorder, which does not create mode mixing, 
while no localization appears in common semiclassical models. We expect
localization to be indeed present in practical situations with 
prevalent 1-D disorder, and the existing experimental evidence appears to 
be consistent with such a prediction.
\end{abstract}
\maketitle 
In the study of low-dimensional devices, suppression of shot noise with 
respect to the value predicted (for the case of a Poissonian noise process)
by Schottky's theorem has represented one of the most active fields of 
investigation in the last two decades. Such a suppression phenomenon,
described by means of the Fano factor, i.e. the
ratio of the actual shot noise power spectral density to the full 
value $2q|I|$ (where $q$ is the value of the elementary charge and $I$ is 
the average value of the current flowing through the device), has been 
predicted and observed in many different mesoscopic structures, and is 
the result of the presence of correlations between charge carriers, which
reduce current fluctuations in the device. From the theoretical point of
view, shot noise suppression has been investigated both with quantum
mechanical and with semiclassical approaches. Such activities have led to the
discovery of ``universal''  values for the Fano factor in specific 
structures;
in particular, for disordered conductors a universal suppression
factor of $1/3$ has been found, both with random matrix theory~\cite{been,jala}
and with a semiclassical approach~\cite{naga}. 
This result has received further confirmation from numerical 
simulations~\cite{kolek,mac} and from experimental evidence~\cite{hen}.

A remarkable addition to these results is a derivation 
by de~Jong and Beenakker~\cite{dejo,dejo2}, who demonstrated that a Fano factor
of $1/3$ is obtained, within a semiclassical model based on the 
Boltzmann-Langevin equation, also for a series of barriers. 
A formulation relying on an equivalent
semiclassical circuit model (similar to that used in \cite{mac2} for a
series of chaotic cavities) leads to the same result.


However, the only existing experimental data~\cite{song}, obtained for 
a GaAs/AlGaAs superlattice, are not in agreement with these semiclassical 
conclusions: they exhibit a Fano factor that depends strongly on barrier 
transmission, in the limit of vanishing applied electric field.

Prompted by this discrepancy, we have performed a quantum calculation of shot 
noise suppression for a structure consisting of cascaded barriers.
While for a conductor with 2-dimensional or 3-dimensional disorder
quantum simulations recover~\cite{kolek,mac} exactly the same 1/3 suppression 
predicted by random matrix theory and by semiclassical models,
this is not the case in the presence of 1-dimensional disorder, i.e. of
randomly spaced cascaded barriers, regardless of the dimensionality of the 
conductor. Our quantum calculations, although performed for a model system,
do, instead, exhibit a behavior consistent with the experimental data.

\begin{figure}
\includegraphics[clip,angle=0,width=8.0cm]{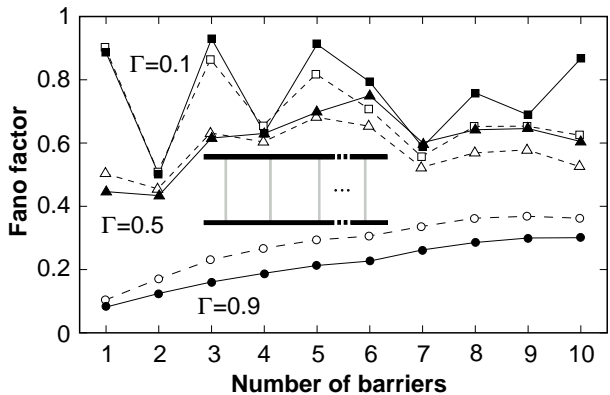}
\caption{
Fano factor for a series of identical barriers, as a function of the number
of barriers for the case of realistic barriers (solid lines), and for
model barriers with a transparency independent of the longitudinal
electron wave vector (dashed lines). Inset: sketch of the analyzed structure.}
\label{fig1}
\end{figure}

In this communication, we focus on the reasons for the discrepancies between 
the semiclassical and the quantum approach, and address the issue of what model 
best represents the situation of a practical experiment.
The structure we have considered (a series of cascaded barriers, sketched
in the inset of Fig.~\ref{fig1} and corresponding to that of 
Refs.~\cite{dejo, dejo2}) can be studied very straightforwardly 
from a numerical point of view. For each fixed value of the longitudinal 
coordinate $x$, the potential profile is constant for $0<y<W$ ($y$ is the
transverse coordinate, and $W$ is the width of the structure: $8~\mu$m in
these calculations), while we assume a hard-wall confinement for $y=0$ and 
$y=W$. Thus, the orthonormal set of transverse modes is the same all over the 
structure, and the tunnel barriers do not introduce any mode-mixing. Since 
the modes are uncoupled, the numerical analysis can be split into a set
of one-dimensional problems, one for each considered mode. 
For the generic $n$-th mode, the S-matrices of an interbarrier region of
length $L$ and of a tunnel barrier of height $U$ and length $l$ are:
\begin{equation}
S_n=
\left(\begin{array}{cc}
0 & \tau_n \\
\tau_n & 0
\end{array}\right)
\,\, ,\quad
S_{B_n}=
\left(\begin{array}{cc}
\rho_{B_n} & \tau_{B_n} \\
\tau_{B_n} & \rho_{B_n}
\end{array}\right) \,\, ,
\end{equation}
where (at the Fermi energy $E_F$)
\begin{eqnarray}
\tau_n = \exp(i k_n L)\,\, ,\quad \rho_{B_n} =
i\,\frac{k_{B_n}^2-k_n^2}{2 k_n k_{B_n}}\,\sin(k_{B_n} l)\,\tau_{B_n}\,\, , 
\nonumber\\
\tau_{B_n} = \left[ \cos(k_{B_n} l)
-i\,\frac{k_n^2+k_{B_n}^2}{2 k_n k_{B_n}}\,\sin(k_{B_n} l) \right]^{-1}
\,\, ,\qquad
\end{eqnarray}
with: $k_n=\sqrt{k_F^2-k_{T_n}^2}$\ ,
$k_{B_n}=\sqrt{k_F^2-k_{T_n}^2-k_U^2}$\ ,\hfill\break
$k_F=\sqrt{2mE_F}/\hbar$\ ,
$k_{T_n}=n\pi/W$ and $k_U=\sqrt{2mU}/\hbar$\ .\hfill\break
For each mode, the scattering matrices of adjacent slices are recursively
composed to find the overall S-matrix and in particular one of its elements,
the transmission $t_n$ of the $n$-th mode through the device.
The conductance and the shot noise power spectral density are then computed 
using the relations~\cite{leso,butt}
\begin{equation}
G=\frac{2\,q^2}{h}\,\sum_n T_n \,\, ,\quad
S_I=4\,\frac{q^3}{h}\,|V|\,\sum_n T_n\,(1-T_n) \,\, ,
\end{equation}
where the sums are performed over all the $N$ modes propagating in the
interbarrier regions, $T_n=|t_n|^2$ and $V$ is the externally applied voltage.
Therefore the Fano factor $\gamma$ can be computed as
\begin{equation}
\gamma=\frac{S_I}{2q|I|}=\frac{\sum_n T_n(1-T_n)}{\sum_n T_n}\,\,\, .
\end{equation}
Before computing the ratio, the values of the numerator and of the denominator
are uniformly averaged over the range of energy $qV$, assuming it much greater 
than $k\theta$ (where $k$ is the Boltzmann constant and $\theta$ is the 
absolute temperature). In particular, our simulations have been performed
using 500 values of energy in a range of $40~\mu$eV around 9.03~meV.

In Fig.~\ref{fig1} we show, with solid lines, the values of the Fano factor 
obtained for an 
$8~\mu$m wide structure made up of a series of identical barriers.
In particular, the reported results are relative to 0.425~nm wide barriers, 
with heights equal to: 0.8~eV (squares), 0.25~eV (triangles) and 0.07~eV 
(circles).
These barriers (at the considered Fermi energy) have an average
transparency $\Gamma=\sum_{n=1}^N |\tau_{B_n}|^2/N$ equal to about 
0.1, 0.5 and 0.9, respectively.
In these simulations the distance between adjacent barriers has been assumed to
be $D+\delta$, where $D=3$~$\mu$m, and (from left to right)
$\delta= 0, 10, -6, 3, -3, -9, 5, -10, -2$~nm.

We see that, as expected, for all transparencies the Fano factor for a single 
barrier is about $1-\Gamma$ (not exactly $1-\Gamma$ because different modes 
experience different transparencies and $\Gamma$ is only an average value). 
For 2 barriers our results are still in agreement with the semiclassical
model of Ref.~\cite{dejo} and, specifically, with the results of its Eq.~(17).
For more than 2 barriers we notice a sharp divergence from the semiclassical
prediction and that no asymptotic 1/3 value is reached. 
Indeed, our results show some dependence on the interbarrier spacings, but 
the overall behavior is already captured by the plots of Fig.~\ref{fig1}. 
A marked difference is observed only in the case of equidistant barriers, 
in which strong resonances between the different interbarrier regions play 
a major role, a case that we do not address in detail in this letter.


\begin{figure}
\includegraphics[clip,angle=0,width=8.0cm]{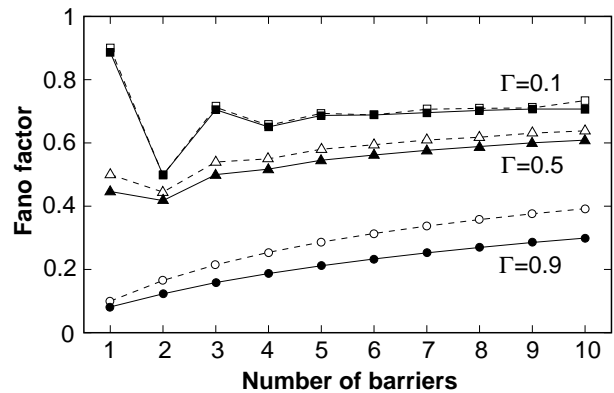}
\caption{Same quantities as in Fig.~\ref{fig1}, but averaged over 50 sets of
interbarrier distances.}
\label{fig2}
\end{figure}

We have also repeated our simulations making the same simplification 
adopted (for analytical convenience) in Ref.~\cite{dejo}, i.e. assuming a
barrier transparency independent of the orthogonal wave vector of the
impinging particle. 
In detail, we have replaced the previously indicated scattering matrix for a 
barrier with 
that of an artificial barrier in which $\rho_{B_n}=-i \sqrt{1-\Gamma}$ and 
$\tau_{B_n}=\sqrt{\Gamma}$, with $\Gamma$ being the wave vector-independent
transparency. No significant variation is observed in the Fano factor when 
such a change is included in our calculation (see dashed lines in 
Fig.~\ref{fig1}).  

In order to remove the dependence of our results on the actual choice of the 
set of lengths of the interbarrier regions, we have performed an average over
several sets~\cite{philos}. It has been shown
that a similar approach is able to reproduce many effects of dephasing on
transport~\cite{pala}.
The results obtained by averaging over energy values and over 50 different
sets of interbarrier distances are shown in Fig.~\ref{fig2} as a function of 
the number of cascaded barriers, for the same transparency values as in
Fig.~\ref{fig1}, assuming either a realistic barrier model (solid lines) or
a wave vector-independent transparency (dashed lines). Also in this case there 
is no clear convergence to a common value of 1/3.

If we consider a situation with cascaded identical barriers
characterized by transparencies that are independent of the wave vector,
and average over random phases,
all propagating modes give the same contribution to the
noise behavior, because the different values of the longitudinal wave vectors
are made ininfluent.
Therefore it becomes possible to perform an analytical
calculation of the Fano factor by considering a single mode 
and integrating over the phase of
$\tau_n$ between $0$ and $2\pi$, for each interbarrier region.
The analytical treatment can be carried out for the cases of 2 and 3
cascaded barriers (with transparency $\Gamma$), for which we obtain,
respectively:
\begin{equation}
\gamma_2=\frac{2(1-\Gamma)}{(2-\Gamma)^2}\quad \textrm{and} \quad
\gamma_3=\frac{3(4-8\Gamma+5\Gamma^2-\Gamma^3)}{16-24\Gamma+9\Gamma^2}
\end{equation}
which are in agreement with the numerical results represented with dashed
lines in Fig.~\ref{fig2}. Coherently with our previous discussion, the result
for 2 barriers coincides with that from the semiclassical model of
Ref.~\cite{dejo}, while that for 3 barriers does not. This is consistent
with the conclusions by F\"orster {\sl et al.}~\cite{foerst}, who show
that only in the case of a single probe (in our structure a probe should
be included between each pair of barriers), current fluctuation statistics
do not depend on the nature (phase averaging, elastic dephasing, or
inelastic) of the probe itself.

The key difference between the semiclassical and the quantum model consists
in the fact that a semiclassical model (unless very peculiar assumptions are 
made~\cite{schanz}) lacks localization as a result of complete incoherence, 
while a
quantum model does exhibit strong localization~\cite{anderson}. 
In particular the
1-dimensional nature of the disorder represented by the randomly placed
barriers makes the system effectively 1-dimensional, regardless of its
actual dimensionality. In this case no mode
mixing is introduced, and therefore the localization length is of the order
of the mean free path; localization occurs beyond this length. 
Instead, in the case of 2- or
3-dimensional disorder, as in Ref.~\cite{mac,kolek}, strong mode mixing makes
the localization length approximately equal to the mean free path times the
number of propagating modes~\cite{tamura}.

In addition, in the absence of mode-mixing, it is not possible to 
consider the interbarrier regions as quasi-reservoirs, 
characterized by a well-defined occupancy, that depends only on the energy. 
This assumption is at the basis of the calculations of 
Refs.~\cite{dejo,dejo2,mac2}, as well as of the semiclassical Monte Carlo
numerical simulation by 
Liu {\em et al.}~\cite{liu}.

Instead, in the absence of mode mixing, only a mode-dependent occupancy  
can be defined; we have computed it for the same structure as in 
Fig.~\ref{fig1}, in which 
all possible electron states can be divided into two sets:
those injected from the left lead and those injected from the right 
lead. Therefore, if we define as $\psi_{n_L}$ and $\psi_{n_R}$ the electron 
wave functions in the generic interbarrier region $\Omega$ resulting from 
an injection of the $n$-th mode (with unit probability current) from the left
or right lead (respectively), the occupancy $f_{n_\Omega}$ in the region 
$\Omega$ for the $n$-th mode can be expressed as the ratio of the partial
density of states related to injection from the left to the total density
of states for that mode~\cite{ianna}:
\begin{equation}
f_{n_\Omega}=\frac{\int_\Omega \left|\psi_{n_L}\right|^2 \,dx \,dy}
{\int_\Omega \left|\psi_{n_L}\right|^2 \,dx \,dy+
\int_\Omega \left|\psi_{n_R}\right|^2 \,dx \,dy}\,\,\, .\label{occu}
\end{equation}

\begin{figure}
\includegraphics[clip,angle=0,width=8.3cm]{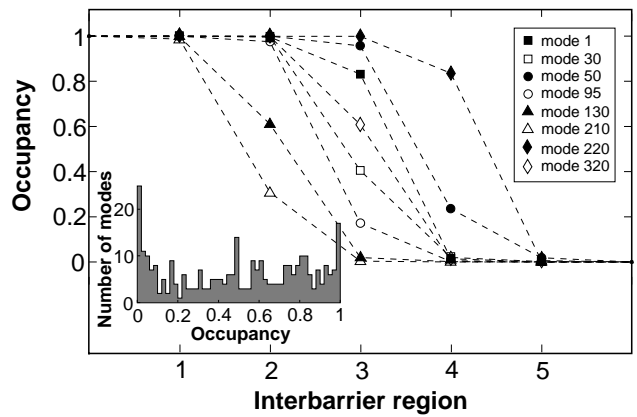}
\caption{Values of the occupancy for 8 modes in the 5 interbarrier 
regions of a series of six unequally spaced tunnel barriers with
$\Gamma=0.1$. Inset: distribution of the occupancy for the propagating 
modes in the 3rd interbarrier region.}
\label{fig3}
\end{figure}

Results for the occupancy in the 5 interbarrier regions of a series of 6 
unequally spaced barriers with an average transparency $\Gamma=0.1$ are 
reported in Fig.~\ref{fig3} for a selection of 8 of the 320 propagating 
modes.  It is apparent that these occupancies assume quite different
values, with a strong dispersion that clearly appears in the inset, 
where we present the distribution of the occupancy
in the region between the 3rd and the 4th barrier. Therefore the
assumption of quasi-reservoir behavior of the interbarrier regions is
definitely not valid in this case.
An exception is confirmed for the case of just two barriers
(thus with a single interbarrier region), in which the occupancies are
all equal and corresponding to the value predicted by semiclassical models.

In the presence of mode-mixing, instead, the localization length $L_l$
is approximately equal, as already mentioned, to the product of the elastic 
mean free path $L_0$
by the number of propagating modes $N$; therefore there can be a range of
device length $L_d$ values in which the condition for diffusive transport
($L_0 \ll L_d \ll N L_0$) is satisfied, and thus the Fano factor
can possibly reach the value 1/3 (as in the case of 2-d or 3-d disorder).  


\begin{figure}
\includegraphics[clip,angle=0,width=8.6cm]{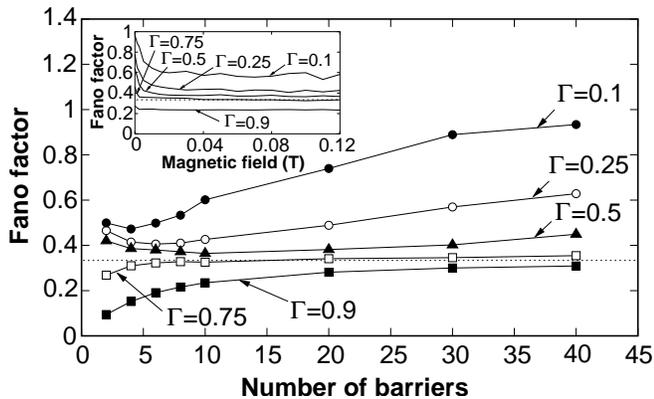}
\caption{Values of the Fano factor obtained for a series of identical
barriers in a 1$\mu$m wide structure with an orthogonal magnetic field
$B=0.1$~T, represented as a function of the number of barriers. In the
inset: Fano factor for a series of 10 identical barriers, as a function
of magnetic field. The dotted lines indicate the diffusive limit of 1/3.}
\label{fig4}
\end{figure}

An adjustable amount of mode-mixing can be introduced by applying 
a magnetic field orthogonal to the plane containing the device.
We have computed the Fano factor for a 1~$\mu$m wide structure with a series
of 10 unevenly spaced barriers, with average interbarrier distance 500~nm.
The results, obtained averaging over a number of interbarrier distance sets
(40 for $\Gamma=0.1$, 30 for $\Gamma=0.25$, 20 for $\Gamma=0.5$, 10 for 
$\Gamma=0.75$ and $\Gamma=0.9$)
are reported in the inset of Fig.~\ref{fig4}, for different choices of the
barrier transparency $\Gamma$.
In detail, the barriers are 66~meV high, with a thickness of 
4, 2.6, 1.56, 0.85 and 0.4~nm for the five considered transparencies;
the Fermi energy is 9.03~meV.
We observe that, after a quick drop, as the magnetic field
increases the Fano factor settles around values that depend on
barrier transparency.  
In the main panel of Fig.~\ref{fig4} we report the Fano factor for a constant
magnetic field of 0.1~T as a function of the number of barriers. We notice
that for large values of the transparency, and thus large values of the mean
free path, a diffusive transport regime (with a Fano factor of 1/3) is 
achieved only for a length much larger than the mean free path. 
On the other hand, for low barrier transparencies, and therefore reduced
mean free path, localization effects appear before reaching the diffusive 
limit. To prevent this, we should significantly increase the number of 
propagating modes and thus the localization length.

The issue is then whether in a practical system 
containing cascaded barriers large enough mode mixing takes place. 
Besides magnetic field, possible mechanisms leading to mode mixing are 
scattering with irregularities in the potential (2-d or 3-d disorder) or 
phonon scattering. 
Scattering due to a disordered potential landscape can well lead to full 
mode mixing, but in such a case the Fano factor of 1/3 characteristic of 
diffusive transport is achieved anyway, independent of the presence
of the barriers and cannot therefore be specifically attributed 
to their action. As far as phonon scattering is concerned, it can  
in principle introduce  mode mixing, but in the presence of strong phonon 
interaction the transport regime would not be the one we are interested in, 
and thermal noise would prevail.

A Fano factor of 1/3 might in principle also be recovered, irrespective of
the degree of mode-mixing, in the presence of a hypothetical elastic mechanism 
capable of suppressing phase coherence completely.

The relatively large values of the phase coherence length 
that can be achieved in modern materials at low temperature and the 
low-field results for the Fano factor presented in Ref.~\cite{song} 
lead us to the conclusion that 
a superlattice or a series of electrostatically 
defined barriers in a channel containing a high-mobility 2DEG  
are more likely to exhibit localization and a Fano factor 
as predicted by our model, rather than a diffusive behavior.  
Numerical approaches along the lines we have presented could be 
instrumental in designing further experimental tests, which should be 
performed on
structures with unevenly spaced barriers.

We are indebted to Prof. C.~W.~J.~Beenakker for useful discussion.


\begin{thebibliography}{999}

\bibitem{been} C.~W.~J.~Beenakker and M.~B\"uttiker, 
Phys.~Rev.~B~{\bf 46}, 1889 (1992).

\bibitem{jala} R.~A.~Jalabert, J.-L.~Pichard, and C.~W.~J.~Beenakker,
Europhys. Lett.~{\bf 27}, 255 (1994).

\bibitem{naga} K.~E.~Nagaev, Phys.~Lett.~A~{\bf 169}, 103 (1992).

\bibitem{kolek} A. Kolek, A. W. Stadler, and G. Halda\'s,
Phys.~Rev.~B~{\bf 64}, 075202 (2001).

\bibitem{mac} M.~Macucci, G.~Iannaccone, G.~Basso, and B.~Pellegrini,
Phys.~Rev.~B~{\bf 67}, 115339 (2003).

\bibitem{hen} M.~Henny, S.~Oberholzer, C.~Strunk, and C.~Sch\"onenberger,
Phys.~Rev.~B~{\bf 59}, 2871 (1999).

\bibitem{dejo} M.~J.~M.~de~Jong and C.~W.~J.~Beenakker, 
Phys.~Rev.~B~{\bf 51}, 16867 (1995).

\bibitem{dejo2} M.~J.~M.~de~Jong and C.~W.~J.~Beenakker, 
Physica~A~{\bf 230}, 219 (1996).


\bibitem{mac2} M.~Macucci, P.~Marconcini, and G.~Iannaccone, 
Int.~J.~Circ.~Theor.~App.~{\bf 35}, 295 (2007).

\bibitem{song}  W.~Song, A.~K.~M.~Newaz, J.~K.~Son, and E.~E.~Mendez,
 Phys.~Rev.~Lett.~{\bf 96}, 126803 (2006).

\bibitem{leso} G.~B.~Lesovik, Pis'ma~Zh.~\'Eksp.~Teor.~Fiz.~{\bf 49}, 513 
(1989) [JETP~Lett.~{\bf 49}, 592 (1989)].

\bibitem{butt} M.~B\"uttiker, Phys.~Rev.~Lett.~{\bf 65}, 2901 (1990).

\bibitem{philos} R.~Landauer, Philosophical Magazine~{\bf 21}, 863 (1970).

\bibitem{pala} M.~G.~Pala and G.~Iannaccone, 
Phys.~Rev.~B~{\bf 69}, 235304 (2004); Phys.~Rev.~Lett.~{\bf 93}, 256803 (2004).

\bibitem{foerst} H.~F\"orster, P.~Samuelsson, S.~Pilgram, and M.~B\"uttiker,
Phys.~Rev.~B~{\bf 75}, 035340 (2007).

\bibitem{schanz} H.~Schanz and U.~Smilansky, Phys.~Rev.~Lett.~{\bf 84}, 1427
(2000).

\bibitem{anderson} P.~W.~Anderson, D.~J.~Thouless, E.~Abrahams, and 
D.~S.~Fisher, Phys.~Rev.~B~{\bf 22}, 3519 (1980).

\bibitem{tamura} H.~Tamura and T.~Ando, Phys.~Rev.~B~{\bf 44}, 1792 (1991).

\bibitem{liu} R.~Liu, P.~Eastman, and Y.~Yamamoto, 
Solid~State~Commun.~{\bf 102}, 785 (1997).

\bibitem{ianna} G.~Iannaccone, Phys.~Rev.~B~{\bf 51}, 4727 (1995).

\end{thebibliography}
\end{document}